# Realizing Kagome Band Structure in Two-Dimensional Kagome Surface States of $R$V$_6$Sn$_6$ ($R$=Gd, Ho)

Shuting Peng,[1,#] Yulei Han,[2,1,#] Ganesh Pokharel,[3,#] Jianchang Shen,[1] Zeyu Li,[1] Makoto Hashimoto,[4] Donghui Lu,[4] Brenden R. Ortiz,[3] Yang Luo,[1] Houchen Li,[1] Mingyao Guo,[1] Bingqian Wang,[1] Shengtao Cui,[5] Zhe Sun,[5] Zhenhua Qiao,[1,*] Stephen D. Wilson,[3,*] and Junfeng He[1,*]

[1]Hefei National Laboratory for Physical Sciences at the Microscale, Department of Physics and CAS Key Laboratory of Strongly-coupled Quantum Matter Physics, University of Science and Technology of China, Hefei, Anhui 230026, China

[2]Department of Physics, Fuzhou University, Fuzhou, Fujian 350108, China

[3]Materials Department and California Nanosystems Institute, University of California Santa Barbara, Santa Barbara, California 93106, USA

[4]Stanford Synchrotron Radiation Lightsource, SLAC National Accelerator Laboratory, Menlo Park, California 94025, USA

[5]National Synchrotron Radiation Laboratory, University of Science and Technology of China, Hefei, Anhui 230026, China

[#]These authors contributed equally to this work.
*To whom correspondence should be addressed:  J.-F.H. (jfhe@ustc.edu.cn), S.D.W. (stephendwilson@ucsb.edu), Z.H.Q. (qiao@ustc.edu.cn).

We report angle resolved photoemission experiments on a newly discovered family of kagome metals $R$V$_6$Sn$_6$ ($R$=Gd, Ho). Intrinsic bulk states and surface states of the vanadium kagome layer are differentiated from those of other atomic sublattices by the real-space resolution of the measurements with a small beam spot. Characteristic Dirac cone, saddle point, and flat bands of the kagome lattice are observed. Our results establish the two-dimensional (2D) kagome surface states as a new platform to investigate the intrinsic kagome physics.

Kagome lattices have attracted much recent attention as a new platform to explore novel correlated topological states. Electronic instabilities and nontrivial topology have been observed, ranging from density waves [1-3], various forms of magnetic order [4-6], a Chern insulator phase [7-10], to the newly reported topological charge order [11]. In momentum space, the kagome physics manifests itself as characteristic band structures, including a Dirac crossing at K, a saddle point at M and a flat band over the Brillouin zone. While some signatures of the characteristic features have been identified in kagome metals by angle resolved photoemission spectroscopy (ARPES) measurements [4,7,8,12-18], the intrinsic 2D kagome physics is often disrupted or destroyed by the complex interlayer and intralayer interactions in the bulk materials [15-17,19,20]. Experimental efforts have been spent to find materials with relatively large separation between kagome layers, where the bulk states of the material may contain more contribution from the kagome layers themselves [15-17].

Different from the bulk states, surface states of the kagome layer follow the in-plane periodic potential of the kagome lattice but decay exponentially along the out-of-plane direction, thus representing the pure intrinsic physics of the 2D kagome lattice. However, the identification of kagome surface states is technically challenging. First, the crystal should contain a pure kagome lattice without other atoms in the same layer. Second, ARPES measurements with real-space resolution are required to probe photoelectrons only from the kagome layer. Third, high data quality is needed to resolve the kagome surface states.

In this Letter, we report the electronic structure of a newly discovered family of kagome metals $R$V$_6$Sn$_6$ ($R$=Gd, Ho), utilizing termination dependent ARPES measurements with a small beam spot. Bulk states of the material and surface states of the vanadium kagome layer are differentiated from those of other atomic sublattices. A Dirac dispersion is observed on the bulk states, although its associated saddle point is invisible. The complete Dirac crossing and saddle point of kagome lattice are identified on the surface states of the vanadium kagome layer, which are reproduced by density functional theory (DFT) calculations. The flat band is also observed, where the surface states and bulk states are degenerate. On the other hand, an extrinsic flat-band-like feature is seen on the Sn-terminated surface, pointing to the importance of differentiating intrinsic kagome physics from the material specific properties in kagome metals. Our work reveals the 2D kagome surface states and the associated characteristic electronic structures for the first time, thus establishes a new platform to realize and investigate the intrinsic physics of kagome lattice.

Single crystals of $R$V$_6$Sn$_6$ ($R$=Gd, Ho) were synthesized from $R$ (pieces, 99.9%), V (pieces, 99.7%), and Sn (shots, 99.99%) via the flux method. The flux mixtures of $R$, V, Sn were heated to 1125°C, dwelled for 12 hours, and then continuously cooled at 2°C/h. The laboratory-grown single crystals of $R$V$_6$Sn$_6$ were separated from the excess flux through conventional centrifuge technique (see Supplemental Material, Fig. S1 [21]). Samples were cleaved at 30 K in ultrahigh vacuum. ARPES results were obtained at Beamline 5-2 of the Stanford Synchrotron Radiation Lightsource (SSRL) of SLAC National Accelerator Laboratory with a total energy resolution of ~11meV and a base pressure of better than 3 × 10$^{-11}$ torr. Some preliminary tests were performed at our lab-based ARPES system and National Synchrotron Radiation Laboratory, University of Science and Technology of China. First-principles calculations were performed by using the projected augmented-wave method [22] as implemented in the Vienna ab *initio* simulation package (VASP) [23,24]. The exchange-correlation interaction was treated with the generalized gradient approximation of the Perdew-Burke-Ernzerhof type [25]. The Hellmann-Feynman force tolerance, kinetic energy cutoff and energy threshold for convergence were set to be 0.01 eV/Å, 520 eV and 10$^{-6}$ eV, respectively. The Fermi surface and surface states were calculated by using WANNIERTOOLS package [26] after obtaining wannier function based tight-binding Hamiltonian as implemented in the WANNIER90 package [27]. The phonon spectrum calculations were carried out by using the density functional perturbation theory as implemented in the PHONONPY package [28].

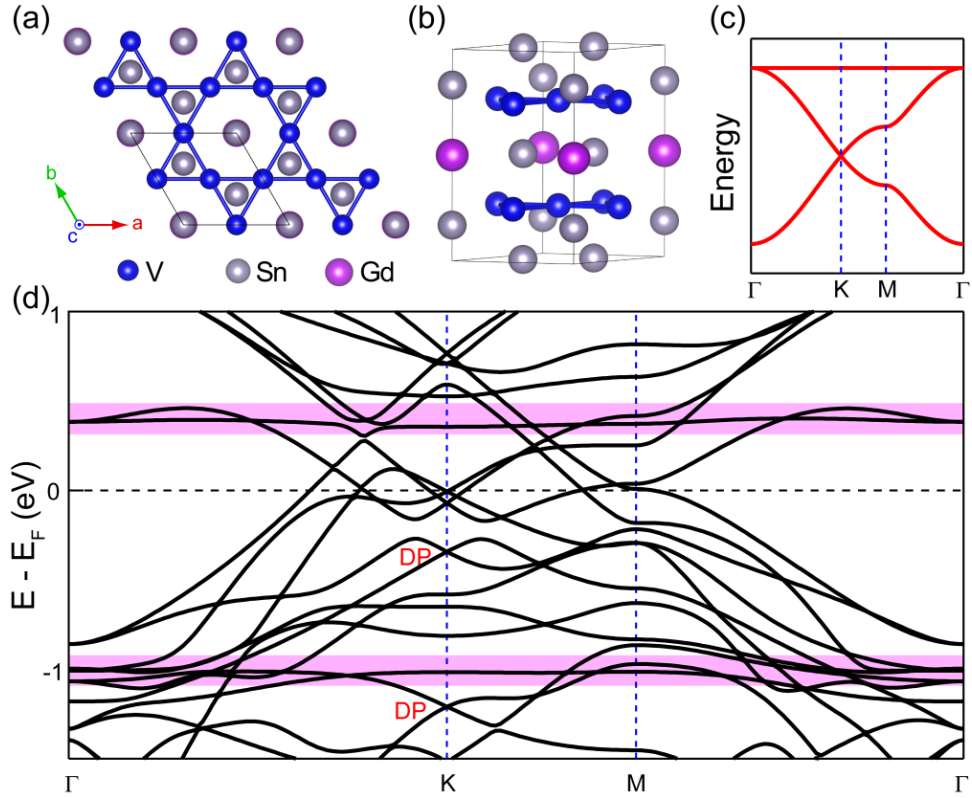

FIG. 1. Crystal structure and calculated band structure of GdV$_6$Sn$_6$. (a),(b) Top view (a) and side view (b) of the GdV$_6$Sn$_6$ crystal structure. V-derived kagome layers are separated by Sn and Gd atomic sublattice layers. (c) Tight-binding calculation of a typical kagome lattice, showing the existence of a flat band, a Dirac crossing at K and a saddle point at M. (d) DFT band structure of GdV$_6$Sn$_6$. The pink shaded regions are an eye-guide for the flat bands. The Dirac points (DP) at K are marked in red.

The crystal structure of $R$V$_6$Sn$_6$ ($R$=Gd, Ho) consists of a V-derived kagome layer with Sn and R atoms distributed in other atomic layers stacked along the out-of-plane direction [Figs. 1(a) and 1(b)], similar to that of $R$Mn$_6$Sn$_6$ ($R$=Gd, Y, Tb, etc.) [29-31]. DFT calculations on GdV$_6$Sn$_6$ exhibit characteristic flat bands and Dirac crossings of a typical kagome network [Figs. 1(c) and 1(d) and Supplemental Material Fig. S2]. Different from its Mn counterpart where complex magnetic structures exist, GdV$_6$Sn$_6$ remains paramagnetic until a very low temperature, which removes the complexity of band reconstruction driven by the magnetic order.

A real-space mapping of the photoelectrons reveals different electronic structures associated with different terminations on the cleaved sample surface of GdV6Sn6 [Fig. 2 and Supplemental Material Fig. S3]. As established in FeSn-based systems, the surface terminations of the sample can be determined by measuring the core levels of Sn [17,32]. While the kagome termination (V layer) only shows bulk states of the Sn 4d core levels [Fig. 2(a) and Supplemental Material Fig. S4], the surface termination with Sn atoms exhibits additional surface peaks near the bulk core levels [Fig. 2(d) and Supplemental Material Fig. S4]. The measured Fermi surface of the kagome termination shows clear resemblance to that of the calculated bulk states of the material [Figs. 3(a) and 3(b)], confirming the identification of the terminations. On the other hand, the Fermi surface of the Sn termination is similar to that of the calculations considering surface states from the sublattice atomic layers with Sn (see

Supplemental Material Fig. S5 [21]). A comparative examination of the band structure on different terminations [Figs. 2(c) and 2(f)] reveals more bands on the Sn termination, being compatible with the existence of surface bands from the Sn-terminated layers. A nearly flat band is observed at ~1eV below the Fermi level ($E_F$) on both terminations, marked by the red dashed line. This is consistent with the DFT calculations [Fig. 1(d)] where a flat band is identified at the same binding energy. In addition, a nearly flat-band-like feature also appears at ~1.1eV below $E_F$ on the Sn termination, mainly between K and M in the momentum space [marked by the yellow dashed line in Fig. 2(f)]. This feature is absent in the band structure of the kagome layer as well as the DFT calculation. Therefore, it doesn't represent the real flat band of the kagome lattice. This observation points to the importance of scrutinizing the intrinsic kagome physics from various flat-band-like features observed on different kagome metals.

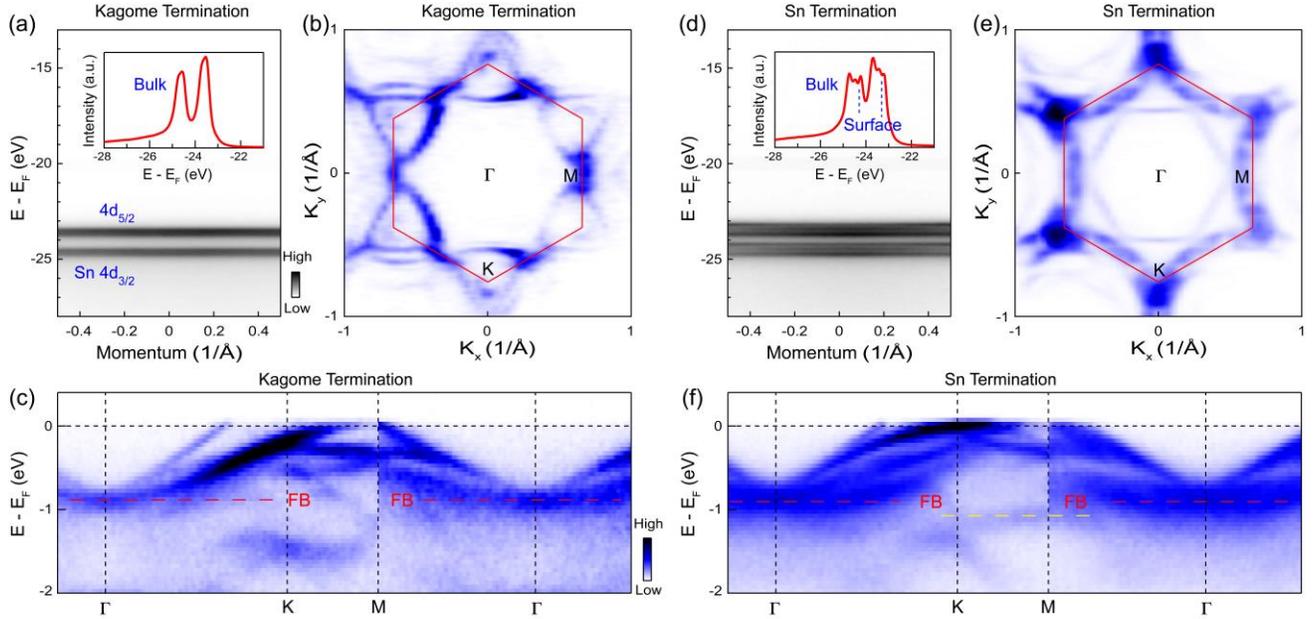

FIG. 2. Termination dependent measurements of the flat band. (a)-(c) Spectrum of the Sn 4d core levels (a), Fermi surface mapping (b), and photoelectron intensity plot of the band structure measured on the kagome termination (c). The inset in (a) shows the integrated energy distribution curve (EDC) of the core levels. The red dashed lines in (c) mark the nearly flat band (FB) similar to that in DFT calculations. (d)-(f) Same as (a)-(c), but for the measurements on the Sn termination. The multiple Sn surface peaks in the core level measurements (d) are associated with the different local environments for Sn atoms with or without adjacent Gd atoms. The yellow dashed line in (f) marks the extrinsic FB-like feature on the Sn termination.

Thus, we focus on the intrinsic electronic structure of the kagome layer hereafter. As shown in Fig. 3, the measured band dispersion near $E_F$ bears clear resemblance to the calculated bulk states of the material. Dirac-like dispersion is observed near the K point of the Brillouin Zone [Fig. 3(g)]. This can also be seen in constant energy maps [Figs. 3(c)-3(e)], where a Fermi pocket around K point shrinks into a bright spot at the Dirac crossing. Photon polarization dependent measurements reveal that different sets of bands in the calculation are selectively enhanced by linear horizontal (LH) and linear vertical (LV) polarized light, respectively [Figs. 3(f)-3(h)]. In particular, the Dirac band near the K point exhibits strong polarization dependence, as shown in Figs. 3(g) and 3(h). This is consistent with the presence of orbital-selective Dirac fermions, reported in several other kagome metals [13,14,16].

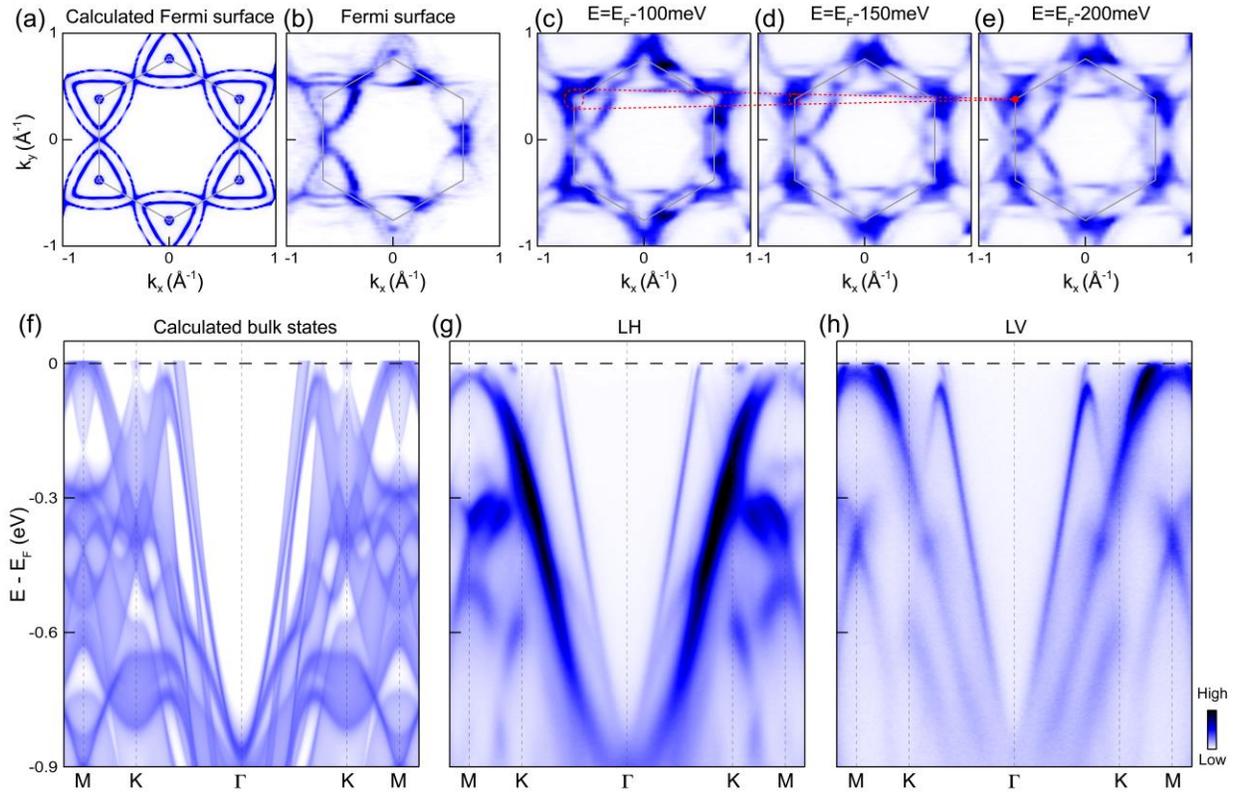

FIG. 3. Electronic structure with Dirac dispersion on the kagome layer of GdV$_6$Sn$_6$. (a) Calculated Fermi surface of the bulk states. (b) Fermi surface measured on the kagome layer with 87eV LH polarized light. (c)-(e) Constant energy maps at -100meV, -150meV and -200meV with respect to $E_F$, respectively. (f) Calculated band structure of the bulk states. (g),(h) Band structure measured with LH (g) and LV (h) polarized light. The red dashed lines in (c)-(e) are an eye-guide for the Dirac dispersion.

A closer examination of the experimental results reveals more electronic structures which are not captured by the calculated bulk states. These features are more clearly seen in the measurements on HoV$_6$Sn$_6$, as shown in Fig. 4 [e.g. the features marked by arrows and dashed lines in Fig. 4(c)]. In order to understand their origin, calculations on the kagome surface layer has been performed [Fig. 4(b) and Supplemental Material Fig. S6]. In addition to the bulk states, four sets of surface state bands are identified [marked by the yellow dashed lines in Fig. 4(b)], which agree well with the additional features observed in the experiment [Fig. 4(c)]. We note that these features are photon energy independent (Supplemental Material Fig. S7 [21]), but are absent on the Sn termination (Supplemental Material Fig. S8). These are consistent with the calculations, further demonstrating that the additional features are not from the bulk states, but associated with the kagome surface states. It is interesting that both a Dirac crossing at K and a saddle point at M are shown in the kagome surface band dispersion [Figs. 4(b)-4(f)], representing the characteristic band structure of the kagome lattice. The absolute binding energies for the surface states are slightly different between the experiment and the calculation. This could be related to the confinement potential on the surface, which is not included in the calculation. A flat band can also be seen at ~1eV below $E_F$, where the kagome surface states and bulk states are largely degenerate (see Supplemental Material Fig. S9).

Next, we discuss the implication of the observations. When distractions from other atomic sublattices are eliminated by termination dependent measurements, the intrinsic Dirac crossing and flat band are observed in the bulk band structure, demonstrating the contribution of kagome layers to the bulk states of the sample. Nevertheless, the characteristic saddle point of kagome lattice is absent in the bulk states (both the calculation and experiment). This is consistent with the earlier reports that the 2D kagome physics is often disrupted or partially destroyed by the complex interactions in the bulk materials [15-17,19,20]. On the other hand, the identification of both the Dirac crossing and saddle point on the kagome surface states points to an ideal realization of the 2D kagome band structure. These results can be naturally understood. The surface states of the kagome layer follow the in-plane periodic potential of the kagome lattice but decay exponentially along the out-of-plane direction. As such, the band structure of the kagome surface states should follow that of the 2D kagome lattice.

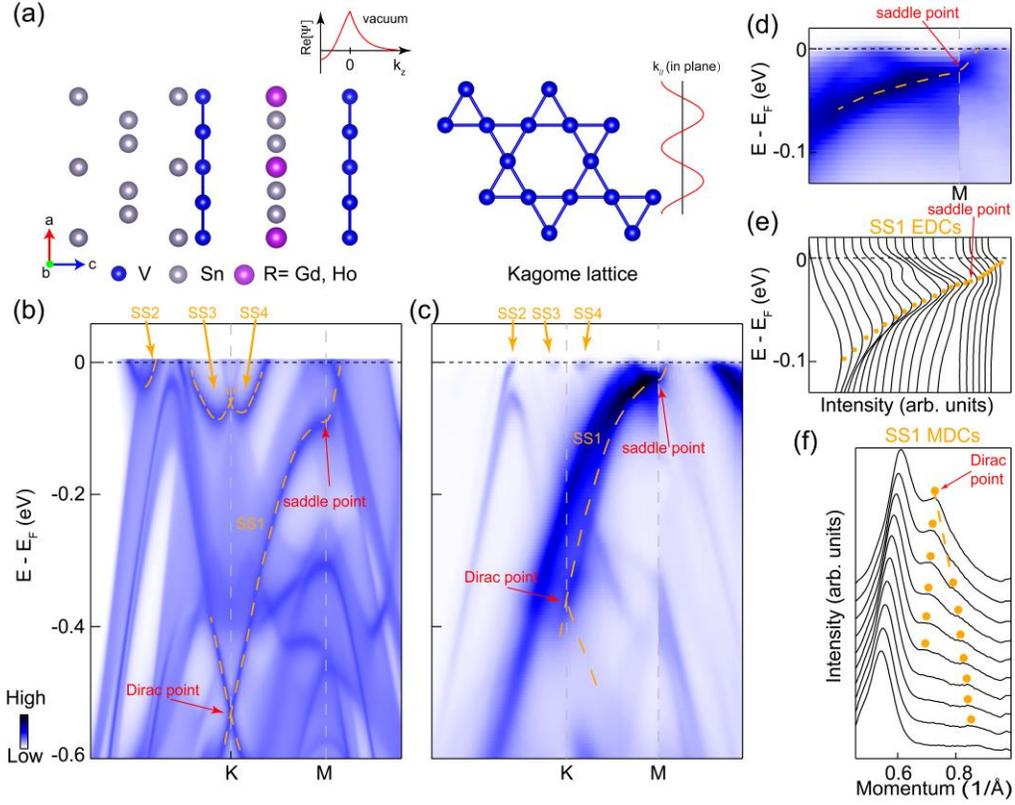

FIG. 4. Kagome surface states of HoV$_6$Sn$_6$. (a) Schematic of the kagome surface layer and kagome surface states. (b) Calculated band structure of the kagome surface layer. (c) Band structure measured on the kagome layer with 60eV LH polarized light. The kagome surface states in (b),(c) are labeled as SS1-SS4, and indicated by the orange arrows and dashed lines. (d) Band structure near M; same as (c) but in an expanded scale to show the saddle point. (e) Raw EDCs near M. (f) Raw momentum distribution curves (MDC) near the Dirac point.

Besides the perfect realization of kagome electronic structure via surface states, we would like to mention the potential manifestation of kagome physics in a bosonic channel of this material. In principle, kagome lattice in bosonic systems can also give rise to flat bands and Dirac cones in the boson dispersion, paralleling that in the electron dispersion [33-37]. Our first principles calculations have mapped out the phonon bands in $R$V$_6$Sn$_6$ (Supplemental Material Fig. S10 [21]). Among them, two flat bands are identified, which arise from the kagome lattice. It would be interesting to explore whether the phonon flat bands couple to the kagome surface states of this material system.

In summary, our termination selective photoemission measurements reveal the intrinsic bulk states and kagome surface states of $R$V$_6$Sn$_6$ ($R$=Gd, Ho). Characteristic electronic structures of the kagome lattice are identified on the kagome surface states. These observations provide an ideal platform to investigate the undisrupted kagome physics in kagome metals.

The work at university of science and technology of China (USTC) was supported by the USTC start-up fund, NNSFC (No. 11974327 and No. 12004369), Fundamental Research Funds for the Central Universities (WK3510000010, WK3510000008 and WK2030020032), and Anhui Initiative in Quantum Information Technologies. We also thank the supercomputing service of AM-HPC and the Supercomputing Center of University of Science and Technology of China for providing the high performance computing resources. S.D.W. and G.P. acknowledge support from the University of California Santa Barbara Quantum Foundry, funded by the National Science Foundation (NSF DMR-1906325). B. R. O. also acknowledges support from the California NanoSystems Institute through the Elings Fellowship program. Use of the Stanford Synchrotron Radiation Lightsource, SLAC National Accelerator Laboratory, is supported by the U.S. Department of Energy, Office of Science, Office of Basic Energy Sciences under Contract No. DE-AC02-76SF00515.